\documentclass[12pt,preprint]{aastex}

\begin{document} 
\title{FEEDBACK REGULATED STAR FORMATION: IMPLICATIONS FOR THE KENNICUTT-SCHMIDT LAW}

\author{Sami Dib\altaffilmark{1}}

\altaffiltext {1} {Astrophysics Group, Blackett Laboratory, Imperial College London, London, SW7 2AZ, United Kingdom; s.dib@imperial.ac.uk}

\begin{abstract} 

We derive a metallicity dependent relation between the surface density of the star formation rate ($\Sigma_{SFR}$) and the gas surface density ($\Sigma_{g}$) in a feedback regulated model of star formation in galactic disks. In this model, star formation occurs in gravitationally bound protocluster clumps embedded in larger giant molecular clouds with the protocluster clump mass function following a power law function with a slope of  $-2.$ Metallicity dependent feedback is generated by the winds of OB stars ($M \gtrsim 5$ M$_{\odot}$) that form in the clumps. The quenching of star formation in clumps of decreasing metallicity occurs at later epochs due to weaker wind luminosities, thus resulting in higher final star formation efficiencies ($SFE_{exp}$). By combining $SFE_{exp}$ with the timescales on which gas expulsion occurs, we derive the metallicity dependent  star formation rate per unit time in this model as a function of $\Sigma_{g}$. This is combined with the molecular gas fraction in order to derive the global dependence of $\Sigma_{SFR}$ on $\Sigma_{g}$. The model reproduces very well the observed star formation laws extending from low gas surface densities up to the starburst regime. Furthermore, our results show a dependence of $\Sigma_{SFR}$ on metallicity over the entire range of gas surface densities in contrast to other models, and can also explain part of the scatter in the observations. We provide a tabulated form of the star formation laws that can be easily incorporated into numerical simulations or semi-analytical models of galaxy formation and evolution.
    
\end{abstract} 


\section{INTRODUCTION}\label{motiv}

The rate at which galaxies convert gas into stars (the star formation rate, SFR) determines their evolution and a wide range of their observable properties. Schmidt (1959) suggested the existence of a relation between the volume density of the SFR and the gas volume density. Kennicutt (1989) derived the surface density of the SFR ($\Sigma_{SFR}$) from H$\alpha$ measurements and combined it with total gas surface density ($\Sigma_{g}$) measurements using HI and CO lines observations in what is now commonly referred to as the Kennicutt-Schmidt  (KS) relation $\Sigma_{SFR} \propto \Sigma_{g}^{n}$. Determining the value of  $n$ has been the subject of intense efforts. In high surface density regions in which massive star formation is present, Kennicutt (1989) found $n=1.3 \pm 0.3$.  Buat et al. (1989) used the UV emission at 2000 $\AA$ of 28 late-type galaxies as a tracer of their recent star formation and found $n=1.65 \pm 0.16$. Kennicutt (1998) included on the same figure data points for normal galaxies and starburst galaxies and obtained $n=1.4 \pm 0.15$. Murgia et al (2002) used radio continuum luminosities as a proxy for the SFR for a sample of 180 galaxies and obtained $\Sigma_{SFR} \propto \Sigma_{H_{2}}^{1.3 \pm 0.1}$, where $\Sigma_{H_{2}}$ is the molecular hydrogen surface density. Martin \& Kennicutt (2001), Komugi et al. (2005), and Schuster et al. (2007) studied the dependence of the radially averaged $\Sigma_{SFR}$ on the gas density profiles in galaxies and found values of $n$ in the range $\sim 1.5 \pm 0.3$, while a few other studies found larger or smaller values (Kuno et al. 1995; Boissier et al. 2003). Heyer et al. (2004) found that in M33, the star formation law follows $\Sigma_{SFR} \propto \Sigma_{g}^{3.3 \pm 0.07}$ but also that $\Sigma_{SFR} \propto \Sigma_{H_{2}}^{1.36 \pm 0.08}$. Zhang et al. (2001) analysed the region-by-region dependence of the $\Sigma_{SFR}$ on $\Sigma_{g}$ in the Antennae and found $n \sim {1.4}$, and Kennicutt et al. (2007) found similar results in M51. Other studies have found linear relations between $\Sigma_{SFR}$ and $\Sigma_{H_{2}}$ or the surface density of molecules that trace higher density gas such as HCN (e.g., Rownd \& Young 1999; Wong \& Blitz 2002;  Gao \& Solomon 2004). Aside from the data of Kennicutt (1998) which included high surface density starburst galaxies, most of the other studies explored surface density ranges extending from a few M$_{\odot}$ pc$^{-2}$ to a few tens of M$_{\odot}$ pc$^{-2}$ (Buat et al. 1989) or up to a few hundreds of M$_{\odot}$ pc$^{-2}$ (Murgia et al. 2002). Heiner et al. (2010) presented a new method in which the volume number densities in the gas clouds surrounding OB associations were determined using a model which considers the atomic hydrogen as a photodissociation product on the clouds surfaces and the UV luminosities were used as a proxy for the SFR. The latter authors obtained an exponent for the Schmidt law of $1.4 \pm 0.2$. Bigiel et al. (2008) combined GALEX ultraviolet and Spitzer 24 $\micron$ observations to derive the SFR for a sample of nearby galaxies on scales of $\sim 750$ pc. They derived $\Sigma_{g}$ using the HI observations of the THINGS survey (Walter et al. 2008) and the CO emission by the BIMA survey of nearby galaxies (Helfer et al. 2003) and the HERA CO-Line Extragalactic survey (Leroy et al. 2009). These new results showed a more complex dependence of $\Sigma_{SFR}$ on $\Sigma_{g}$. Bigiel et al. (2008) found that $n \sim 1$ for $\Sigma_{g}$ in the range $9-80$ M$_{\odot}$ pc$^{-2}$ and that the $\Sigma_{SFR}-\Sigma_{g}$ relation has a much steeper slope in the regime of $\Sigma_{g} < 9$ M$_{\odot}$ pc$^{-2}$. 
    
Several ideas have been proposed in order to explain the origin of the KS law. One of the early explored scenarios is one in which stars form as a result of gravitational instabilities in galactic disks over a characteristic timescale which is the local free-fall time of the gas and which is given by $t_{ff,g} \propto \rho_{g}^{-0.5}$, where $\rho_{g}$ is the local gas volume density. For a constant scale height of the disk, $\rho_{g} \propto \Sigma_{g}$ and thus $\Sigma_{SFR} \propto \Sigma_{g}/t_{ff,g} \propto \Sigma_{g}^{1.5}$ (Madore 1977; Li et al. 2006). Wong \& Blitz (2002) argued that the value of the KS law slope is related to the value of the molecular fraction $f_{H_{2}}=\Sigma_{H_{2}}/\Sigma_{g}$ and Blitz \& Rosolowsky (2006) showed that $f_{H_{2}}$ is related to the pressure of the interstellar medium. Tassis (2007) and Wada \& Norman (2007) suggested that the value of $n$ is related to the width of the density probability distribution function of the interstellar gas and to the threshold density that is associated with the gas tracer. The origin of the KS laws has also been extensively investigated using numerical simulations which were able to reproduce KS laws with slopes of $\sim 1.5-2$ (Kravtsov 2003; Tasker \& Bryan 2006; Shetty \& Ostriker 2008; Robertson \& Kravtsov 2008; Schaye \& Dalla Vecchia 2008; Gnedin et al. 2009; Papadopoulos \& Pelupessy 2010; Gnedin \& Kravtsov 2011; Hopkins et al. 2011). Escala (2011) pointed out that a correlation exists between the largest mass-scale for structures not stabilised by rotation and the SFR while Abelardo Zamora-Aviles \& V\'{a}zquez-Semadeni (2011) argued that the SFR in gravitationally collapsing molecular clouds is regulated by feedback from massive stars. Krumholz et al. (2009a, KMT09) proposed a model for the star formation laws in galaxies which combines: a) a prescription for the molecular gas fraction as a function of $\Sigma_{g}$, b) a description of the behaviour of Giant Molecular Clouds (GMCs) as a function of $\Sigma_{g}$, and c) a model, based on turbulence, which describes the conversion of a fraction of the GMCs mass into stars per unit time. KMT09 computed several star formation laws for various values of the interstellar gas metallicities and compared their results to a compilation of observational results which included the Bigiel et al. (2008) data at low $\Sigma_{g}$ and up to the starburst regime data of Kennicutt (1998). In this work we propose that the fraction of dense molecular gas that is converted into stars per unit time is mainly determined by metallicity dependent feedback in protocluster forming regions which are themselves embedded in larger GMCs. We explore the effects of this metallicity dependent prescription on the star formation laws from low surface density regions up to the starburst regime and find an excellent agreement with the observations. 

\section{THE FEEDBACK REGULATED AND METALLICITY DEPENDENT STAR FORMATION LAW}

In the KMT09 model, the surface density of star formation $\Sigma_{SFR}$ is given by:

\begin{equation} 
\Sigma_{SFR}=\Sigma_{g}~f_{H_{2}} \frac{SFE_{ff}}{t_{ff}},
\label{eq1}
\end{equation} 

\noindent where $\Sigma_{g}$ is the total gas surface density, and $SFE_{ff}$ is the dimensionless star formation efficiency and which corresponds to the mass fraction of the molecular gas that is converted into stars per free-fall time $t_{ff}$ of the GMCs in which stars form. The  $SFE_{ff}$ that is used in KMT09 is the one derived by Krumholz \& Mckee (2005) based on a model which describes the gravo-turbulent regulation of star formation in GMCs. The quantity $f_{H_{2}}$ is the mass fraction of the total gas that is in molecular form. Krumholz et al. (2009b) have shown that a good approximation of $f_{H_{2}}$ in a given atomic-molecular complex is given by:

\begin{equation}
f_{H_{2}} (\Sigma_{comp},Z^{'}) \approx 1-\left[ 1+ \left(\frac{3}{4} \frac{s}{1+\delta} \right)^{-5}\right]^{-1/5},
\label{eq2}
\end{equation}

\noindent where $s={\rm ln} (1+0.6~\chi)/(0.04\Sigma_{comp} ({\rm M_{\odot} pc^{-2}})~Z^{'})$, $\chi=0.77 (1+3.1~Z^{'0.365}$), $\delta=0.0712~(0.1~s^{-1}+0.675)^{-2.8}$, and $Z^{'}$ is the metallicity in units of the solar value. As pointed out by KMT09, $\Sigma_{comp}$ refers to the surface density of atomic-molecular complexes of typical scales of $\sim 100$ pc whereas the typical current spatial resolution on which $\Sigma_{g}$ is measured is several hundred pc or larger. Thus, it is appropriate to consider that $\Sigma_{comp}= c~\Sigma_{g}$ where $c \geq 1$ is a clumping factor which approaches unity as the spatial resolution of the observations approaches 100 pc. The only dependence of $\Sigma_{SFR}$ on metallicity in their model is through the term $f_{H_{2}}$. Their model also assumes that stars form in a distributed way in GMCs with an efficiency per unit time that depends only on their dynamical properties (their virial parameter and the Mach number). They also assume that the characteristic GMC mass is determined by the Jeans mass in the galaxy and it relates to $\Sigma_{g}$ by:

\begin{equation} 
M_{GMC}=37 \times 10^{6} \left(\frac{\Sigma_{g}}  {85~\rm{M_{\odot}~pc^{-2}}} \right) {\rm M_{\odot}}. 
\label{eq3}
\end{equation} 
 
Here, we propose a model for star formation which describes the formation of stars in dense protocluster forming clumps which are themselves embedded in larger GMCs. Star formation in these clumps is regulated by stellar winds feedback from the newly formed OB stars (of masses $\gtrsim 5$ M$_{\odot}$). In this model, dense cores form in the clump with a given core formation efficiency per free-fall time of the clump ($CFE_{ff}$). The level and driving scale of turbulence and the strength of the magnetic fields influence the value of the $CFE_{ff}$. Dib et al. (2010a) showed that the $CFE_{ff}$ in a clump/cloud can vary from $\sim 0.06$ to $0.33$ when the mass-to-magnetic flux ratio changes from $\mu_{B}=2.2$ to $8.8$ (in units of the critical value for collapse). This strong dependence on the strength of the magnetic field shows that a regulation of the star formation efficiency which is based solely on turbulence as postulated by KMT09 is not entirely satisfactory. Dib et al. (2011) varied the value of the $CFE_{ff}$ between $0.1$ and $0.3$ to account for variations due to different levels of turbulence and magnetic field strengths in the clump. The cores have lifetimes of a few times their free-fall time after which they are turned into stars with a one core-to-one star efficiency conversion of 1/3. Once the cumulated effective kinetic energy from the stellar winds of the newly formed OB stars is larger than the gravitational binding energy of the clump, gas is expelled from the clump and subsequent core and star formation are quenched. Dib et al. (2011) measured the final star formation efficiency, $SFE_{exp}$, when gas is expelled from the protocluster clump and found that the value of $SFE_{exp}$ depends weakly on the value of the $CFE_{ff}$ but shows a strong dependence on metallicity. This dependence on metallicity is related to the strong dependence of the wind power on metallicity. Stronger winds (i.e., associated with higher metallicities) will result in the gas being expelled from the protocluster clump at an earlier epoch and thus reduce the value of $SFE_{exp}$. The relevant mass-scale in our model is the characteristic mass of the protocluster clumps in the GMCs and not the characteristic mass of the GMCs as in KMT09. The characteristic clump mass is given by:

\begin{equation}
 M_{char}= \int_{M_{cl,min}}^{max(M_{cl,max},M_{GMC})} M_{cl} N(M_{cl}) dM_{cl}, 
 \label{eq4}
\end{equation}

\noindent where $M_{GMC}$ is given by Eq.~\ref{eq3}. $M_{cl,min}$ is the minimum mass of a protocluster clump which we take to be $2.5 \times 10^{3}$ M$_{\odot}$ (this guarantees, for final SFEs  in the range of 0.05-0.3 a minimum mass for the stellar cluster of $\sim 50$ M$_{\odot}$). There might be a dependence of $M_{cl,max}$ on the GMC mass but for simplicity we adopt everywhere $M_{cl,max}=10^{8}$ M$_{\odot}$ since the most massive known clusters have masses of the order of a few $10^{7}$ M$_{\odot}$ (Walcher et al. 2005, Portegies Zwart et al. 2010). $N (M_{cl})$ is the mass function of protocluster forming clumps which we take to be $N (M_{cl})=A_{cl} M_{cl}^{-2}$ in agreement with the results of Dib et al. (2011) and Parmentier (2011), and $A_{cl}$ is a normalisation coefficient given by $A_{cl} \int_{M_{cl,min}}^{max(M_{cl,max},M_{GMC})} N(M_{cl}) dM_{cl}= \epsilon$, where $0 < \epsilon < 1$ is the mass fraction of the GMCs that is in protocluster clumps at any given time. In this work we use $\epsilon=0.5$. Fig.~\ref{fig3} (top) displays $M_{char}$ as a function of $\Sigma_{g}$. Based on the above described model, we propose the following star formation law: 
 
\begin{equation} 
\Sigma_{SFR}=\Sigma_{g}~f_{H2} \frac{\left<SFE_{exp}\right>} {\left<t_{exp}\right>},
\label{eq5}
\end{equation}  

\noindent where $\left<SFE_{exp}\right>$ and $\left<t_{exp}\right>$ are, respectively, the characteristic SFE and the epoch at which gas is expelled from the protocluster region for the clump mass distribution associated with a given $\Sigma_{g}$. Writing $\left<t_{exp}\right>$ in terms of the characteristic clump free-fall time $\left<t_{ff}\right>$ ($n_{exp}=t_{exp}/t_{ff}$), Eq.~\ref{eq2} becomes:

\begin{eqnarray}
\Sigma_{SFR}=\Sigma_{g}~f_{H_{2}} \frac{\left<SFE_{exp}\right>} {\left<n_{exp}\right>} \frac{1}{\left<t_{ff}\right>} \nonumber \\
                                 ~=\Sigma_{g}~f_{H_{2}} \frac{\left<f_{\star,ff}\right>}{\left<t_{ff}\right>}.
\label{eq6}                  
\end{eqnarray}

\begin{figure}
\plotone{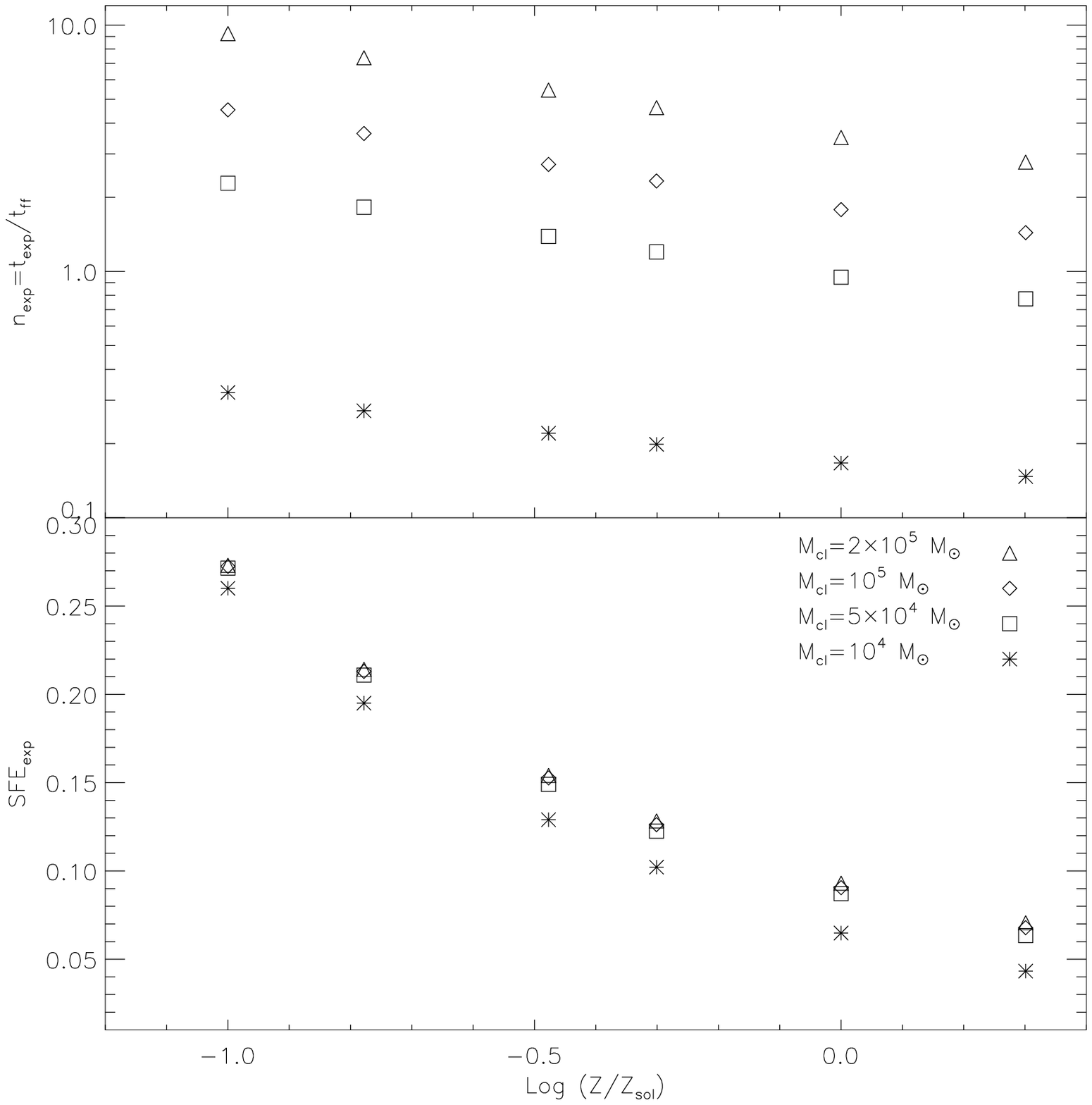}
\vspace{1.5cm}
\caption{Dependence of the quantities $SFE_{exp}$ (final star formation efficiency) and $n_{exp}=t_{exp/t_{ff}}$ (ratio of the expulsion time to the free-fall time) for selected values of the protocluster forming clump masses and metallicities. These results are based on the models of Dib et al. (2011).}
\label{fig1}
\end{figure}

\begin{figure}
\plotone{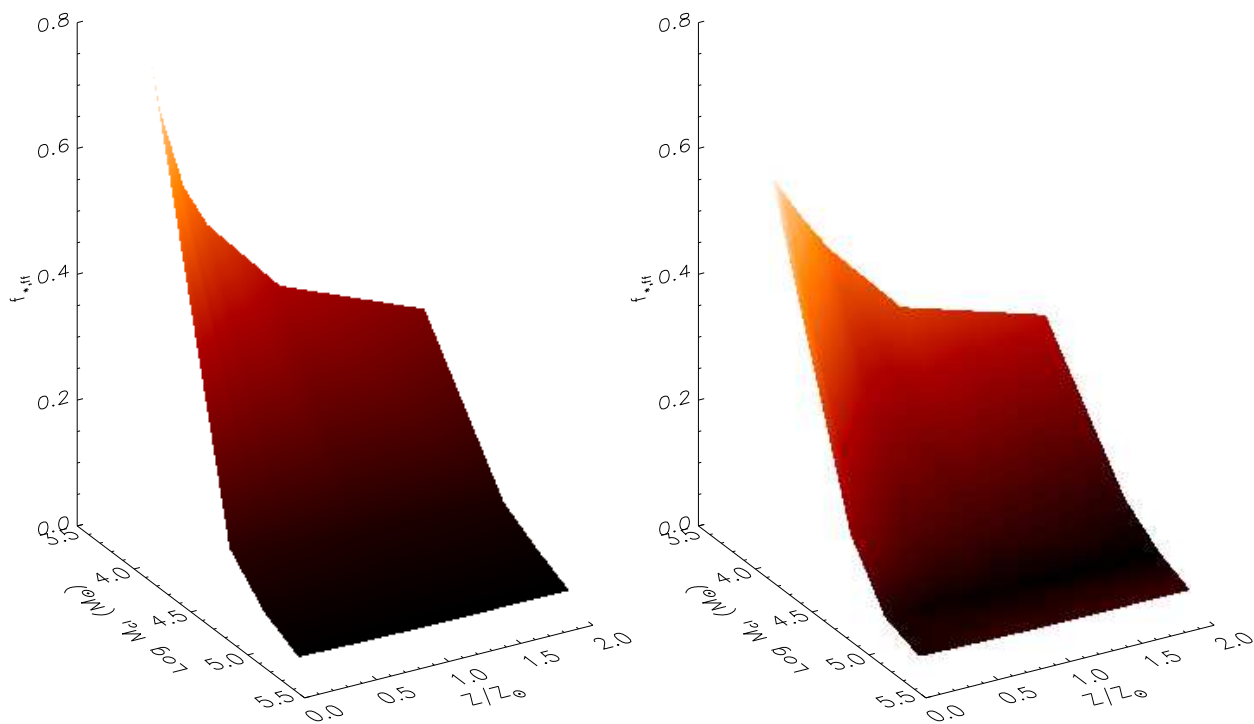}
\caption{Star formation efficiency per unit free-fall time in the protocluster clump in the metallicity-dependent feedback model of Dib et al. (2011). The model uses a core-to-star efficiency conversion factor of 1/3. The left panel displays $f_{\star,ff}$ as a function of both $M_{cl}$ and $Z^{'}=Z/Z_{\odot}$ in the original data. The right panel displays the analytical fit function to this data set (Eq.~\ref{eq7}).} 
\label{fig2}
\end{figure}

\begin{figure}
\plotone{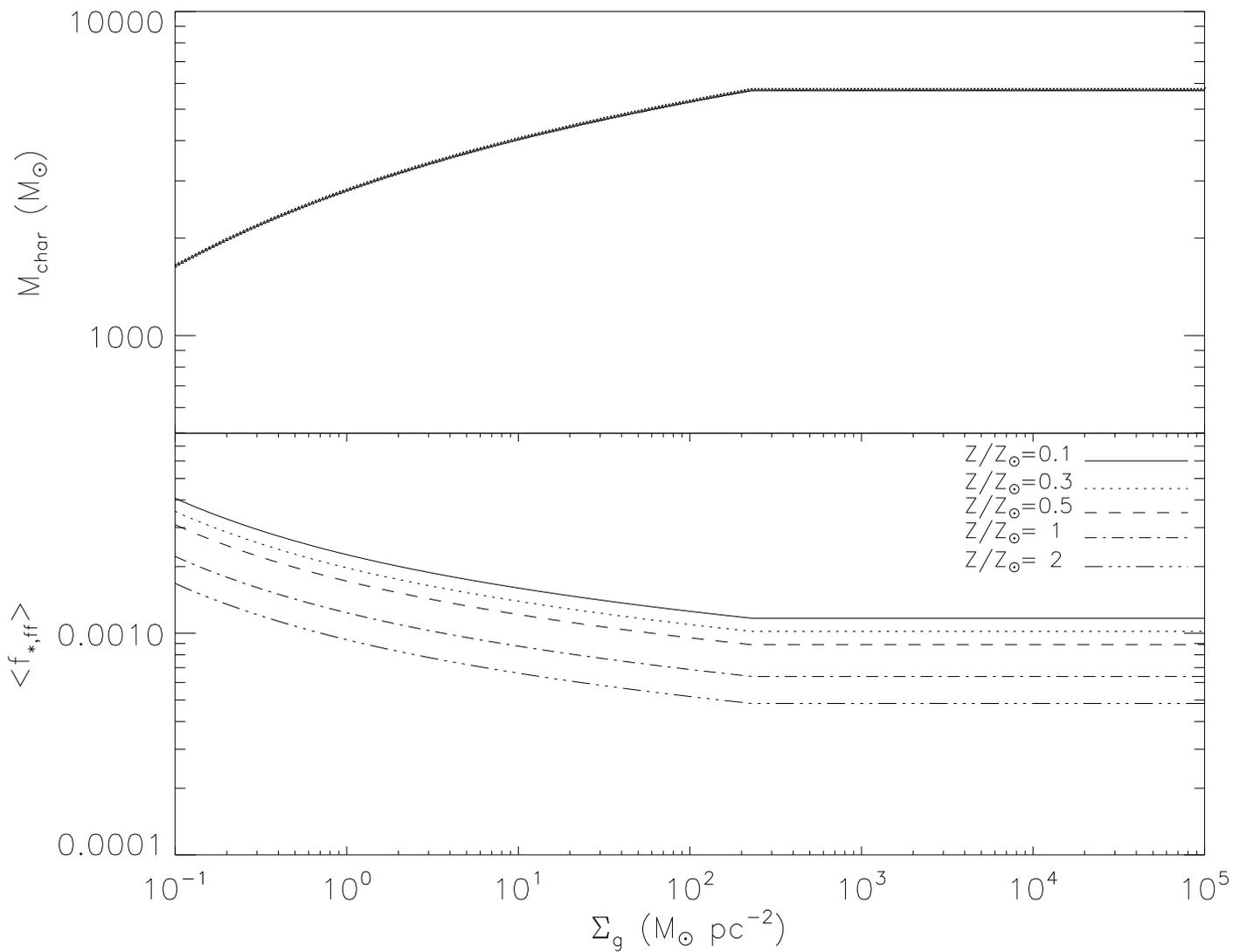}
\vspace{1.5cm}
\caption{Characteristic clump mass as a function of the gas surface density (Eq.~\ref{eq4}, top panel) and the star formation efficiency per unit free-fall time in this feedback regulated model of star formation (Eq.~\ref{eq8}).}
\label{fig3}
\end{figure}

In Eq.~\ref{eq6}, $\left<f_{\star,ff}\right>$ represents the characteristic star formation efficiency per free-fall time of the clump mass distribution in the feedback regulated star formation model. In order to calculate $\left< f_{\star,ff} \right>$ we use the results of Dib et al. (2011). Fig.~\ref{fig1} displays the dependence of $SFE_{exp}$ and $t_{exp}$ on clump mass and metallicity for a serie of models in which the $CFE_{ff}=0.2$. The $SFE_{exp}$ depends stronly on metallicity but weakly on mass whereas $t_{exp}$ displays a clear dependence on both quantities. The quantity $f_{\star,ff}=SFE_{exp}/n_{exp}$ is displayed in Fig.~\ref{eq2} as a function of mass and metallicity (left panel). A fit to the $f_{\star,ff}$ ($M_{cl},Z^{'}$) data points with a 2-variables second order polynomial yields the following relation (shown in Fig.~\ref{fig2}, right panel):

\begin{eqnarray}
f_{\star,ff}(M_{cl},Z^{'})=  11.31-4.31{\rm log(M_{cl})} + 0.41 {\rm [log(M_{cl})]^{2}}  \nonumber \\ 
           - 8.28 Z^{'} + 3.20 Z^{'} {\rm log (M_{cl})} - 0.32 Z^{'} {\rm [log(M_{cl})]^{2}}  \nonumber \\
          + 2.30 Z^{'2} - 0.89 Z^{'2} {\rm log(M_{cl})} + 0.08 Z^{'2} {\rm [log(M_{cl})]^{2}}.
\label{eq7}
\end{eqnarray} 

Using Eq.~\ref{eq7}, it is then possible to calculate $\left<f_{\star,ff}\right>$:  

\begin{equation}
\left<f_{\star,ff}\right> (Z^{'}, \Sigma_{g}) = \int_{M_{cl,min}}^{max(M_{cl,max},M_{GMC})} f_{\star,ff} (M_{cl},Z^{'}) N(M_{cl}) dM_{cl}. 
\label{eq8}
\end{equation}

\begin{figure}
\plotone{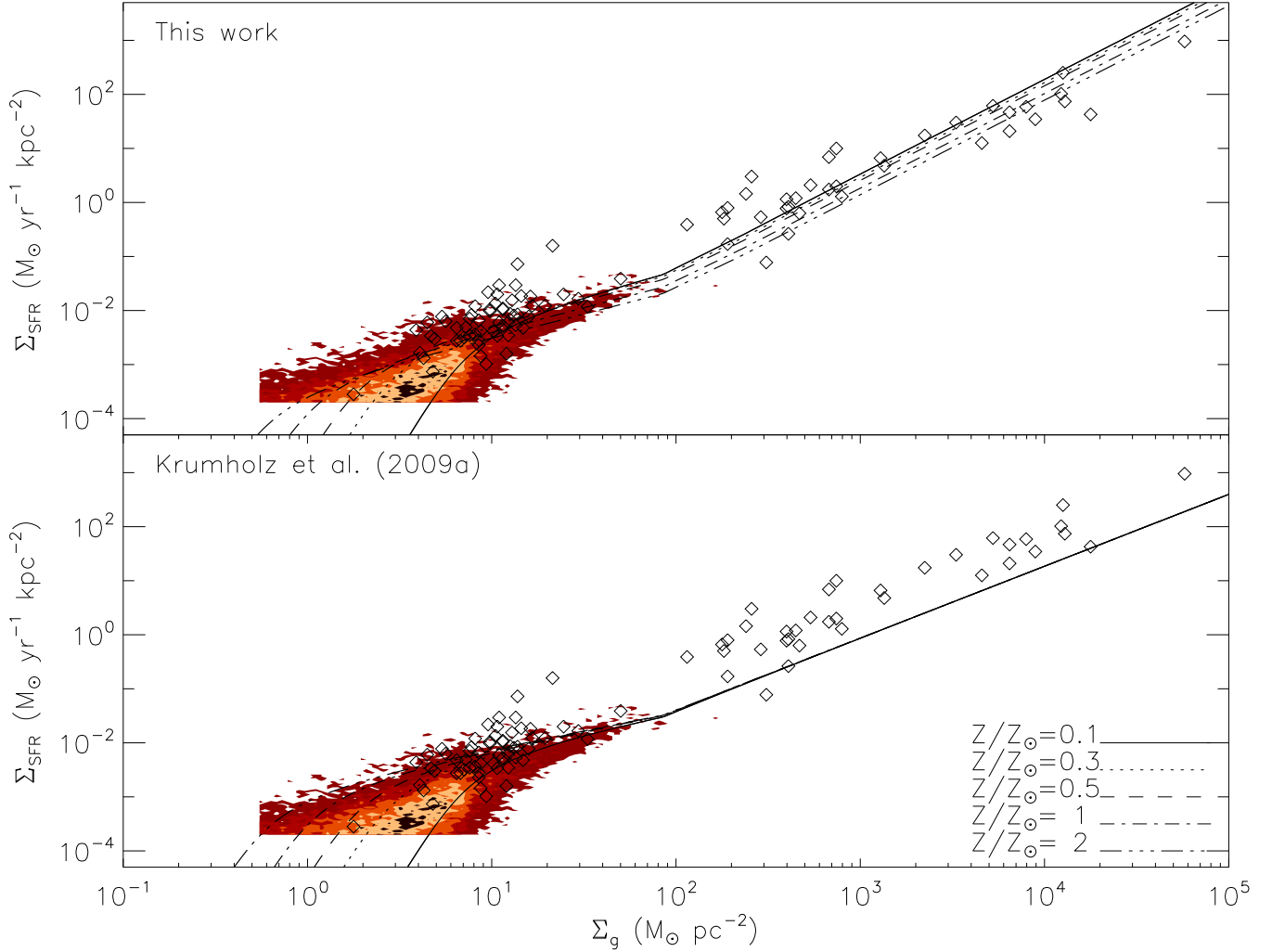}
\vspace{1.5cm}
\caption{Star formation laws in the feedback-regulated star formation model (this work, top panel), and in the Krumholz et al. (2009a) model (bottom panel). Overplotted to the models are the normal and starburst galaxies data of Kennicutt (1998) and the combined sub-kpc data (4478 subregions) for 11 nearby galaxies from Bigiel et al. (2008,2010). The Bigiel et al. data is shown in the form of a 2D histogram with the colour coding corresponding, from the lighter to the darker colours to the 1,5,10,20, and 30 contour levels. The displayed theoretical models cover the metallicity range $Z^{'}=Z/Z_{\odot}=[0.1,2]$.}
\label{fig4}
\end{figure}

Fig.~\ref{fig3} (bottom) displays $\left<f_{\star,ff}\right>$ ($Z^{'},\Sigma_{g}$) for values of $Z^{'}$ in the range [$0.1-2$]. As in KMT09, we also assume that there is a critical value of $\Sigma_{g}$ below which clumps are pressurized by their internal stellar feedback, and for the sake of comparison, we adopt the same value of $\Sigma_{g,crit}=85$ M$_{\odot}$ pc$^{-2}$ such that $\Sigma_{cl}=\Sigma_{g,crit}$ where $\Sigma_{g} < \Sigma_{g,crit}$ and $\Sigma_{cl}=\Sigma_{GMC}=\Sigma_{g}$ when $\Sigma_{g} \geq \Sigma_{g,crit}$. In Eq.~\ref{eq6} $\left<t_{ff}\right>$ can be approximated by the free-fall time of the clump with the characteristic mass $t_{ff} (M_{char})=8 \Sigma_{cl}^{' -3/4} M_{char,6}^{1/4}$ Myr where $M_{char,6}=M_{char}/10^{6}$ M$_{\odot}$. We also adopt the same $f_{H_{2}}$ as in KMT09 which is given by Eq.~\ref{eq2}. With the above elements, the star formation law can be re-written as:

\begin{eqnarray}
\Sigma_{SFR} &=& \frac{8} {10^{6}} f_{H_{2}} (\Sigma_{g},c,Z^{'}) \Sigma_{g}  \nonumber \\
  \times & & \left \{\begin{array} {cc}  \frac{\left<f_{\star,ff}\right> (Z^{'},\Sigma_{g})} {M_{char,6}^{1/4}(\Sigma_{g})} &  ; \frac{\Sigma_{g}} {85~{\rm M_{\odot}~pc^{-2}}} < 1\\
  \frac{\left<f_{\star,ff}\right> (Z^{'},\Sigma_{g})} {M_{char,6}^{1/4}(\Sigma_{g})} \left(\frac{\Sigma_{g}}{85~{\rm M_{\odot} pc^{-2}}}\right)^{3/4}  & ; \frac{\Sigma_{g}}{85~{\rm M_{\odot}~pc^{-2}}} \geq 1 \end{array} \right \},
\label{eq9}
\end{eqnarray}

\noindent where $\Sigma_{SFR}$ is in M$_{\odot}$ yr$^{-1}$ kpc$^{-2}$, $M_{char}$ is given by Eq.~\ref{eq4}, and $\left<f_{\star,ff}\right>$ by Eqs.~\ref{eq7} and \ref{eq8}. Fig.~\ref{fig4} (top panel) displays the results obtained using Eq.~\ref{eq9} for $\Sigma_{g}$ values starting from low gas surface densities up to the starburst regime. The results are calculated for the metallicity values of $Z^{'}=[0.1,0.3,0.5,1,2]$, and use a value of $c=5$ (the structure of the results is displayed in Tab.~\ref{tab1} and the full set of results is available in the electronic version of the paper). The results are compared to the sub-kpc data of Bigiel et al. (2008) and to the normal and starburst galaxies results of Kennicutt (1998) and also to the KMT09 results (Fig.~\ref{fig4}, bottom panel). Our models fits remarkably well the observational results over the entire range of surface densities. Furthermore, the segregation by metallicity extends beyond the low surface density regime up to the starburst regime where a segregation in metallicity of $\sim 0.3$ dex is observed (dependence is $Z^{',-0.3}$), in contrast to the KMT09 models which do not contain a metallicity dependence in the intermediate to high surface density regimes. Furthermore the solar metallicity curve in our model overlaps with a significant fraction of the sub-regions in the data of Bigiel et al. (2008,2010) in contrast to the KMT09 model. The values of the slopes in the high ($\Sigma_{g} > 85$ M$_{\odot}$ pc$^{-2}$) and intermediate ($10$ M$_{\odot}$ pc$^{-2} < \Sigma_{g} < 85$ M$_{\odot}$ pc$^{-2}$ ) surface density regimes are [1.75,1.74,1.74,1.74,1.74] and [1.20,0.97,0.93,0.90,0.88] for $Z^{'}$=[0.1,0.3,0.5,1,2], respectively. The slope of the $\Sigma_{g}-\Sigma_{SFR}$ relation increases to $\sim 5.65$ at low surface densities ($\Sigma_{g} < 1$ M$_{\odot}$ pc$^{-2}$). A few additional factors may enhance the variations that are due to metallicity. Here, we have derived the star formation law using clump models in which we adopted a galactic averaged value of $CFE_{ff}=0.2$. As shown in Dib et al. (2011), variations around this value in the range $0.1-0.3$ (due to variations in global galactic conditions such as the strength of the magnetic field) as indicated by numerical simulations of (Dib et al. 2010a) can lead to variations by up to $\sim 0.5$ dex in the values of the $SFE_{exp}$ and $t_{exp}$. Furthermore, for the sake of simplicity, the adopted shape of the clump mass function and its minimum and maximum mass cutoffs, as well as the fraction of the GMCs mass present in protocluster clumps have been assumed to be independent of $\Sigma_{g}$. Galaxy-to-galaxy variations in the shape of $N (M_{cl})$, $M_{cl,min}$, $M_{cl,max}$ and $\epsilon$ may add additional scatter to the KS relations. Additionally, Dib et al. (2007) and Dib et al. (2010b) showed that the accretion of gas by the cores and/or core coalescence in the protocluster clumps can modify the shape of the IMF and hence the value of the $SFE_{exp}$. An efficient accretion by the cores in the protocluster clumps as well as variations in the clumps properties that can be attributed to variations in their galactic environments can lead to a steepening of the $(SFE_{exp}-t_{exp})-metallicity$ relations and cause additional scatter in the star formation law. Support to our proposed feedback regulated star formation model may come from the recent interesting results of Shi et al (2011) who showed the existence of an 'extended KS law' which points to the role of the existing stars in regulating the galactic star formation efficiency.          
 
\section{CONCLUSIONS}\label{conc}

In this work, we have derived the star formation law (Kennicutt-Schmidt law, KS) in galaxies using a model which incorporates the following elements: we use a description of the molecular fraction in atomic-molecular complexes. The latter quantity has been derived by Krumholz et al. (2009a, KMT09). As in KMT09, we also assume that there is a critical surface gas density ($\Sigma_{g}=85$ M$_{\odot}$ pc$^{-2}$) above which the protoclusters clump and their parent giant molecular clouds switch from being pressurised from within by stellar feedback to being confined by the external interstellar medium pressure. However, in contrast to the model of KMT09 in which star formation is purely regulated by turbulence, we assume that the bulk of star formation occurs in protocluster forming clumps whose star formation efficiency is regulated by the metallicity dependent stellar winds of OB stars that form in the clump. Using results from the Dib et al. (2011) models on the final star formation efficiency and the duration of the star formation process before gas expulsion, we calculate the metallicity dependent star formation efficiency per unit time in this feedback regulated star formation scenario. The combination of these three elements allows us to construct the star formation law in galaxies going from low gas surface densities up to the starburst regime. Our models exhibit a dependence on metallicity over the entire range of considered gas surface densities and fit remarkably well the observational data of Bigiel et al. (2008) and Kennicutt (1998). This dependence on metallicity of the KS relation may well explain the scatter that is seen in the observationally derived relations. Tabulated results of the star formation laws are provided in electronic form and can be easily incorporated into numerical simulations and analytical models of galaxy formation and evolution.  

\acknowledgements

I thank the referee for useful suggestions and Jakob Walcher, Sergiy Silich, Subhanjoy Mohanty, George Helou, and Frank Bigiel for useful discussions. I acknowledge support from STFC grant ST/H00307X/1
  
{}

\begin{deluxetable}{cccccc}
\tabletypesize{\footnotesize}
\tablecaption{This table displays the surface density of the star formation rate ($\Sigma_{SFR}$) as a function of the gas surface density ($\Sigma_{g}$) for five selected values of  the metallicity ($Z^{'}=Z/Z_{\odot}$)$^a$.}
\tablewidth{0pt}
\tablehead{
\colhead{${\rm log}(\Sigma_{g}$)} & \colhead {$\Sigma_{SFR}$} & \colhead {$\Sigma_{SFR}$} & \colhead {$\Sigma_{SFR}$} & \colhead {$\Sigma_{SFR}$} & \colhead {$\Sigma_{SFR}$} \\
\colhead {}                                          & \colhead {$Z^{'}=0.1$}        & \colhead{$Z^{'}=0.3$} & \colhead{$Z^{'}=0.5$} & \colhead{$Z^{'}=1$} & \colhead{$Z^{'}=2$} \\
\colhead{(M$_{\odot}$ pc$^{-2}$)} & \colhead{(M$_{\odot}$ yr$^{-1}$ kpc$^{-2}$)} &  \colhead{(M$_{\odot}$ yr$^{-1}$ kpc$^{-2}$)} & \colhead{(M$_{\odot}$ yr$^{-1}$ kpc$^{-2}$)} & \colhead{(M$_{\odot}$ yr$^{-1}$ kpc$^{-2}$)}&  \colhead{(M$_{\odot}$ yr$^{-1}$ kpc$^{-2}$)} }
 
\startdata
$-0.50$ &  $6.31\times 10^{-11}$ & $3.98 \times 10^{-9}$ & $2.85 \times 10^{-8}$ & $3.53 \times 10^{-7}$ & $4.31 \times 10^{-6}$ \\
\enddata
\tablecomments{
               (a) The abridged version here is shown to illustrate the table structure. The online machine readable-table contains 276 values of $\Sigma_{SFR}$ for each selected value of the metallicity and for values of ${\rm log}(\Sigma_{g})$ (M$_{\odot}$ pc$^{-2}$) that sample the range $[-0.5,5]$.}
\label{tab1}
\end{deluxetable}

\end{document}